\documentclass[11pt]{amsart}

\hoffset - 1.9 truecm
\usepackage{amsmath,amssymb}
\usepackage{graphicx}
\usepackage{color}

\def\r{\rho}
\def\k{\kappa}
\def\l{\lambda}

\def\vf{\varphi}

\def\W{\Omega}

\def\bq{{\bf q}}

\def\bv{{\bf v}}
\def\bg{{\bf g}}

\def\bn{{\bf n}}
\def\bp{{\bf p}}

\def\bA{{\bf A}}

\def\cP{{\mathcal P}}

\def\bea{\begin{eqnarray*}}
\def\eea{\end{eqnarray*}}
\def\rot{\nabla \times}
\def\div{\nabla \cdot}
\def\R{\mathbb{R}}

\makeatletter

\@addtoreset{equation}{section}

\begin{document}

\title[Superfluid transition in liquid helium]{A thermodynamically consistent Ginzburg-Landau model for superfluid transition in liquid helium}

\author[A. Berti]{Alessia Berti}\address{Facolt\`a di Ingegneria, Universit\`a e-Campus, 22060 Novedrate (CO), Italy} \email{alessia.berti@ing.unibs.it}

\author[V. Berti]{Valeria Berti}\address{Dipartimento di Matematica,
Universit\`a di Bologna, Piazza di Porta S. Donato 5, I-40126
Bologna, Italy} \email{valeria.berti@unibo.it}

\date{}

\begin{abstract}
In this paper we propose a thermodynamically consistent model for superfluid-normal phase transition in liquid helium, accounting for variations of temperature and density. The phase transition is described by means of an order parameter, according to the Ginzburg-Landau theory, emphasizing the analogies between superfluidity and superconductivity. 
The normal component of the velocity is assumed to be compressible and the usual phase diagram of liquid helium is recovered. Moreover, the continuity equation leads to a dependence between density and temperature in agreement with the experimental data.
\end{abstract}

\maketitle

\bigskip


\section*{Introduction}
The phenomenon of superfluidity occurs mainly in liquid helium below a characteristic temperature $\theta_{\lambda}$. Above $\theta_\lambda$, helium behaves like a conventional fluid with small viscosity. However, when the temperature is lowered below $\theta_\lambda$, liquid helium undergoes a phase transition characterized by the ability of the liquid to flow across narrow channels without apparent friction. Helium has two stable isotopes $^4$He and $^3$He that become superfluid at low temperatures. The most common isotope is $^4$He whose transition temperature, called the $\lambda-$point, is about $2.17 K$. The normal phase of $^4$He  is called the He I-phase and the superfluid state is said He II.

In this paper we propose a phenomenological model to describe the phase transition in $^4$He.
The first model to study the behavior of $^4$He was the two-fluid model, suggested by Tisza (\cite{Tisza}) and developed by Landau (\cite{landau}).
According this theory, when the temperature is under $\theta_\l$, each particle of the fluid is endowed with two different excitations at the same instant: one of these is the superfluid velocity, denoted by $\bv_s$, the other one is the normal velocity $\bv_n$. The density of the fluid is the sum of a normal and a superfluid component
$$
\rho=\rho_n+\rho_s
$$ 
and the total current density is given by
$$
{\bf j}=\rho_n\bv_n+\rho_s\bv_s.
$$
If the temperature overcomes the $\l-$point, the density $\rho_s$ vanishes, so that liquid helium becomes a normal fluid.

The two-fluid model has been widely adopted to describe some phenomenological aspects of superfluidity, when the involved velocity of the fluid is quite small (see \cite{TT} and references therein).
Besides the two-fluid model, some authors follow the one-fluid theory of liquid helium, which is based on the extended irreversible thermodynamics (see \cite{LH, M} for instance). They analyze the behavior of liquid helium II considered as a unique substance obeying a suitable Navier-Stokes equation.
More recently in \cite{F_super} the author studies the phase transition between helium I and helium II in the framework of the Ginzburg-Landau theory, by considering this passage as a second order phase transition and introducing a scalar variable $\vf$ as order parameter such that $\vf^2$ represents the concentration of the superfluid phase. This point of view emphasizes also the analogies between superfluidity and superconductivity (\cite{Mend,TT}). Indeed, as in the two-fluid model, the velocity of the fluid is due to a normal and a superfluid excitation; however the superfluid component $\bv_s$ is supposed to satisfy an evolution equation similar to the differential equation governing the motion of the superconducting electrons inside a superconductor (\cite{Fab06}).

In our paper we consider a generalization of this model by keeping into account variations of the mass density of the fluid and variations of the temperature. The main assumption, distinguishing our model by the one proposed in \cite{Fab06}, is that the normal component is a compressible fluid. Accordingly, the pressure becomes a new variable of the problem and the divergence of the normal component $\bv_n$ satisfies a constitutive equation depending on the phase variable $\vf$.
When the fluid is in the normal state, the evolution equation for the normal component $\bv_n$ reduces to the Navier-Stokes equation.
\begin{figure}[h]
\begin{center}
\includegraphics[scale=0.3]{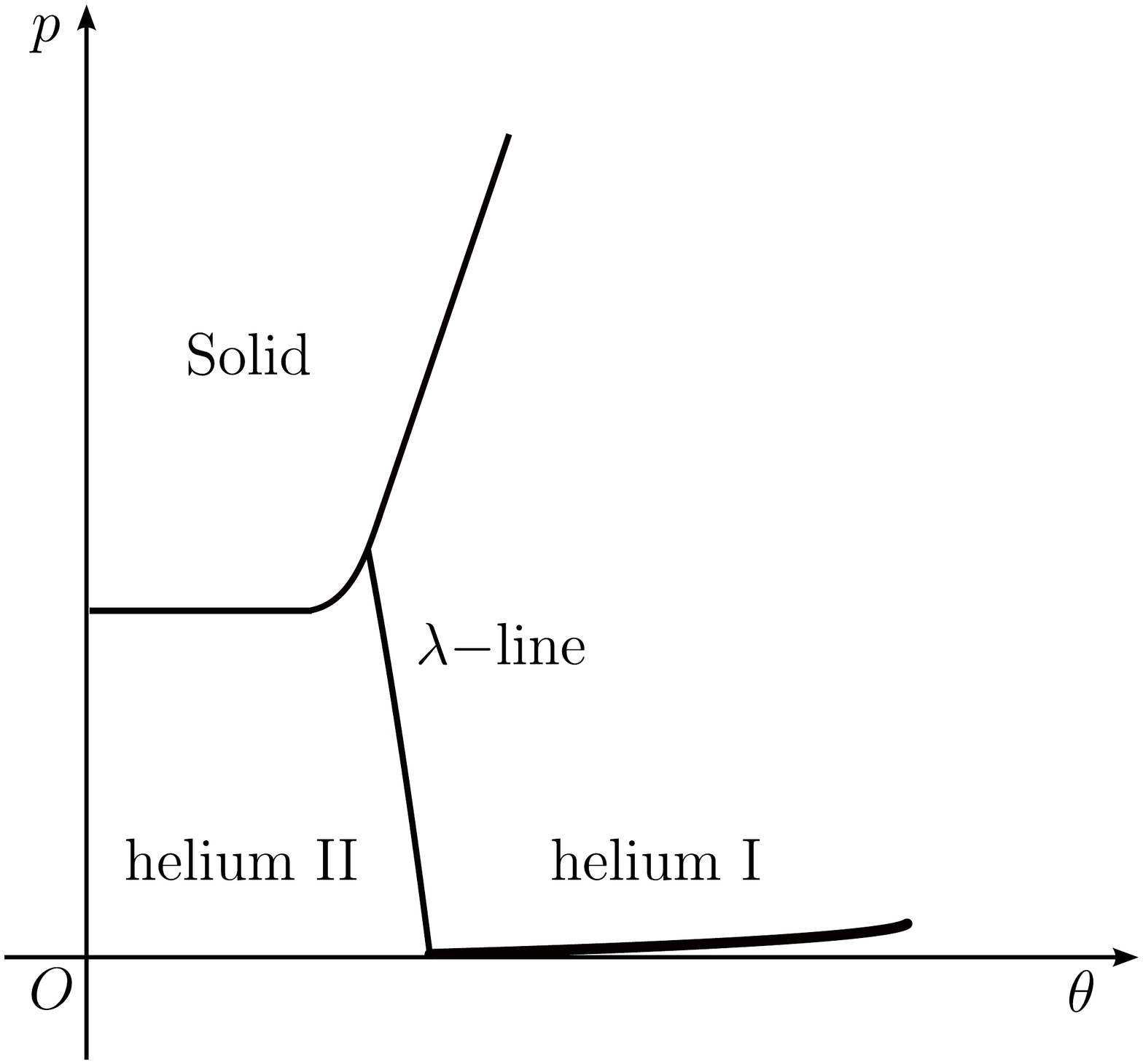}
\caption{Phase diagram of $^4$He.}
\label{lam}
\end{center}
\end{figure}

The occurrence of the pressure in the phase equation allows us to recover in the phase diagram the line ($\l-$line) separating the normal from the superfluid phase (see Fig. \ref{lam}).

We prove that our model is consistent with thermodynamic principles, deducing the differential equation for the temperature from the energy balance law and proving that Clausius-Duhem inequality is satisfied.
In particular we assume that the heat flux is the sum of two contributions: the first term is proportional to the gradient of the temperature (Fourier law), the other one is due to the superfluid transition since it involves the superfluid component $\bv_s$ and the phase variables $\vf$. This point of view is not dissimilar from models based on the extended irreversible thermodynamics, which interpret the  superfluid velocity as a kind of heat flux (\cite{LH, M}).

In the last section of the paper we show that the differential system governing the evolution of the fluid can be written by means of a different set of variables, similar to the unknown fields used in the context of superconductivity (\cite{FM,T}). 
Such a formulation could be useful in proving some analytical results concerning the well-posedness of the system.

\section{The Ginzburg-Landau equation for the superfluid concentration}

As known, the passage from the normal phase to the superfluid state is a second-order phase transition, since no latent heat is involved \cite{Mend}. Therefore we propose a model to describe the phenomenon in the context of the Ginzburg-Landau theory. The first step is the identification of a suitable order parameter characterizing the state of the material. Here we introduce a scalar variable $\vf$ (phase-field), such that $\vf^2$ represents the concentration of the superfluid phase. 
Thus the values of $\vf^2$ are bounded in the interval $[0,1]$ with $\vf^2=0$ in the normal phase and $\vf^2\neq 0$ in the superfluid regime.
The variable $\vf$ provides a measure of the internal order structure of the material, since the superfluid phase is considered a more "ordered" state than the normal one \cite{landau}.
Accordingly, the differential equation governing the evolution of $\vf$ can be interpreted as a balance law on the internal order structure. For a general treatment of balance laws in continuous bodies with microstructure see \cite{Capriz, PPG}.
The interpretation of the Ginzburg-Landau equation as a balance equation has been proposed by Fried and Gurtin who introduce the notion of microforces (\cite{G,fried}). In this context every change of the order parameter is related to the existence of microforces which expend power on the atomic configurations inside the material. A similar approach has been proposed in \cite{Fab06} where a balance of the internal order structure is postulated. The common idea of the two approaches is that, during the transition, in any sub-region $A$ of the body, the power expended on the atoms by the lattice is balanced by the power expended across the boundary $\partial A$ by the configurations external but neighboring to $A$ and by the power expended by sources external to the body. We briefly recall the interpretation of the Ginzburg-Landau equation as a balance law, adopting the terminology of \cite{Fab06}.

Let us consider a superfluid occupying a bounded subset $\Omega \subset \R^3$ with regular boundary $\partial \Omega$, whose outward normal is denoted by $\bn$. 
For any sub-body $A\subset \W$, we denote by
$\mathcal{S}^{i}(A)$ the rate of absorption of the order structure  per unit time, defined as 
\begin{equation}
\mathcal{S}^{i}(A)=\int_{A}\rho kdv,
\end{equation}
where  $\rho$ is the mass density and $k$ is the \textit{internal specific structure order}.
Similarly, the \textit{external order structure} $\mathcal{S}^{e}(A)$
is written in the form 
\begin{equation}
\mathcal{S}^{e}(A)=\int_{\partial A}\mathbf{p}\cdot \mathbf{n}
ds+\int_{A}\rho \sigma dv,
\end{equation}
where the vector $\mathbf{p}$ denotes the \textit{order structure flux} coming from the boundary and $\sigma$ is
the \textit{structure order supply}.

Hence, the order structure balance is expressed 
by the equality 
\begin{equation}
\int_{A}\rho kdv=\int_{\partial A}\mathbf{p}\cdot \mathbf{n}ds+\int_{A}\rho
\sigma dv\qquad\qquad
\forall A\subset\W.  \label{1.3}
\end{equation}

In local form, the integral equality (\ref{1.3}) leads to the equation 
\begin{equation}
\rho k=\nabla \cdot \mathbf{p}+\rho \sigma.  \label{1.4}
\end{equation}

Hence, the quantity $\bp$ can be assimilated to a vector stress and $k$ and $\sigma$ to internal and external microforces distributed in the domain $\W$.
As usual in phase transition problems, we assume $\sigma=0$. Moreover,  the functions $k$ and $\mathbf{p}$ are defined by means of the constitutive equations 
\begin{eqnarray}
k &=& \tau \dot\vf+ \theta_\l \, F^{\prime }(\vf)+mG^{\prime }(\vf)  \label{1.4b}
\\
\mathbf{p}&=&\frac{\r}{\kappa^2}\nabla \vf,  \label{1.4c}
\end{eqnarray}
where the superposed dot stands for the time derivative, $m>0$ is a suitable coefficient depending on the variables that induce the transition, the potentials $F$ and $G$ characterize the order and the feature of the transition and $\k,\tau$ are positive constants. 
Accordingly, the evolution equation for the order parameter reads
\begin{equation}\label{eq_vf0}
\tau \r \dot \vf = \frac{1}{\k^2} \nabla \cdot (\r\nabla \vf) - \r \theta_\l \, F'(\vf) -\r m G'(\vf).
\end{equation}

\begin{center}
\begin{figure}[h]
\includegraphics[scale=0.4]{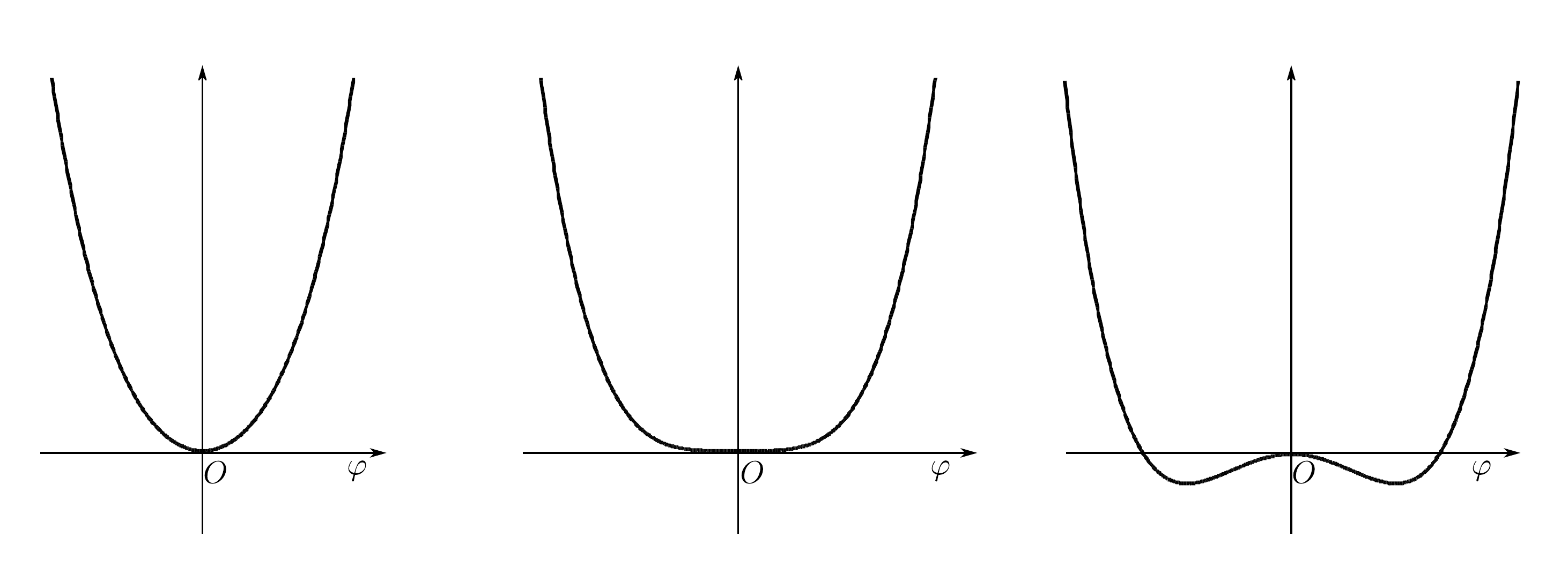}
\caption{Plot of the potential $W$ for different values of $m$: in the left $m>\theta_{\lambda}$, in the middle $m=\theta_{\lambda}$, in the right $m<\theta_{\lambda}$.}
\label{figW}
\end{figure}
\end{center}

In steady and homogeneous conditions ({\it i.e.} $\dot\vf=0$ and $\nabla\vf=0$) the solutions of (\ref{eq_vf0}) are the stationary points of the function
$$
W(\vf)=\theta_\l \, F(\vf)+mG(\vf).
$$
A typical choice adopted for second-order phase transitions is 
\begin{equation}
F(\vf)=\frac{\vf^{4}}{4}-\frac{\vf^{2}}{2},\qquad G(\vf)=\frac{\vf^{2}}{2}.
\label{1.4d}
\end{equation}
Such expressions are the same used in the classical Ginzburg-Landau theory of superconductivity (\cite{GL}).
The functions $F$ and $G$ satisfy the following properties (see Fig. \ref{figW}):
\begin{itemize}
\item[(i)] the function $W$ admits its minimum value at $\vf=0$ when $m\geq \theta_\l$ and at $\vf=\pm\vf_0(m)$, $0<\vf_0(m)<1$, when $0<m<\theta_\l$;
\item[(ii)]  $\vf_0(m)\to 0$ as $m\to \theta_\l$ and  $\vf_0(m)\to1$ as $m\to 0$.
\end{itemize}

As a consequence the transition occurs when $m=\theta_\l$ and $m>\theta_\l$ identifies the normal phase.

Notice that there is no physical distinction between the positive and negative values of $\vf$ since the physical quantity is $\vf^2$. As a consequence, we can consider $\vf \in [0,1]$ as well as $\vf \in [-1,1]$. In the latter case, we have to require that the potentials $F$ and $G$ are even functions.

Finally, we associate to equation (\ref{eq_vf0}) the usual Neumann boundary condition
\begin{equation} \label{bc_vf}
\bp \cdot \bn |_{\partial \Omega} = 0.
\end{equation}

\section{Evolution equation for the velocity}
According to Landau's viewpoint, we assume that each particle of the superfluid exhibits a normal and a superfluid excitation. Thus the velocity $\bv$ is given by the
superposition of such flows. In particular, we let
\begin{equation}\label{v}
\bv = (1-\vf^2)\bv_n + \vf^2 \bv_s.
\end{equation}
where $\bv_n$, $\bv_s$ will be called normal and the superfluid components, respectively.
This does not mean that superfluid is considered as a mixture of two fluids.
Indeed, a particle of the material is endowed with two simultaneous excitations (normal and superfluid) when $\vf\neq 0$, while its velocity coincides with $\bv_n$ when $\vf=0$. 
However, the expression (\ref{v}) of the velocity is not in contrast with the traditional theories of superfluids which assume
$$
\rho\bv=\rho_n\bv_n+\rho_s\bv_s,
$$
when $\rho_n$ and $\rho_s$ are identified respectively with $\rho_n=\rho(1-\vf^2)$ and $\rho_s=\rho\vf^2$.  

The variables $\rho$ and $\bv$ are related by the continuity equation
\begin{equation}\label{eq_rho}
	\dot \r + \r \div\bv=0.
\end{equation}

By paralleling \cite{F_super}, the differential equations governing the evolution of the component $\bv_s$ are
\begin{eqnarray}
\label{eq_vs}
\dot \bv_{s} &=& - \nabla \phi_s - \rot \bv_n -  \r \vf^2\bv_s + \nabla \theta
\\
\label{eq_phi}
\div[ \r \vf^2 \bv_s] &=& -\tau \k^2 \r \vf^2\phi_s
\end{eqnarray}
where  $\phi_s$ is a suitable scalar function referable to a ``pressure'' due to the superfluid component.
It is worth noting that \eqref{eq_vs} and \eqref{eq_phi} are similar to the evolution equations governing the motion of
superconducting electrons \cite{GL}, emphasizing the evident analogies between superfluidity and
superconductivity.

We associate to these equations the boundary condition
\begin{eqnarray}
\label{bc_vs}
\bv_{s} \cdot \bn |_{\partial \Omega} &=& 0.
\end{eqnarray}
A further boundary condition on $\bv_s$ has to be prescribed. Our model allows us to choose such a condition in an arbitrary way, since, as we will see in Sect. 4, the condition (\ref{bc_vs})  is sufficient to ensure the vanishing of the power flux at the boundary of the domain. In the superconductivity model it is assumed 
\begin{eqnarray}
\label{bc_rotvs}
(\rot\bv_s) \times \bn |_{\partial \Omega} &=& \omega,
\end{eqnarray}
where $\omega$ is a known function. Therefore we can assume (\ref{bc_rotvs}) by analogy with that model.

For the velocity $\bv_n$ we propose the following equation
\begin{equation}\label{eq_vn}
	\rho(1-\vf^2)\dot\bv_n = \rot\dot\bv_s + \nu \nabla(\div\bv_n) -\nabla p -\frac{1}{\kappa^2}\div(\nabla\vf\otimes\nabla\vf)-\r \vf^2 \nabla\theta +\r {\bf g},
\end{equation}
where $\nu$ is the viscosity coefficient, $p$ is the pressure and $\bg$ denotes the external force density.
By applying the curl operator to equation \eqref{eq_vs}, we obtain
$$
\rot \dot \bv_{s} = - \rot \rot \bv_n -  \rot[\r \vf^2 \bv_s].
$$
A substitution into equation \eqref{eq_vn} leads to the generalizing Navier-Stokes equation
$$
\rho(1-\vf^2)\dot\bv_n + \rot \rot \bv_n  - \nu \nabla(\div\bv_n) = -\nabla p -\frac{1}{\kappa^2}\div(\nabla\vf\otimes\nabla\vf)-\r \vf^2 \nabla\theta - \rot[ \r \vf^2\bv_s] + \rho \bg.
$$
We append to \eqref{eq_vn} the usual boundary condition:
\begin{equation}
	\label{bc_vn}
\bv_n |_{\partial \Omega} = {\bf 0}.
\end{equation}

%================================================================================
%================================================================================
%================================================================================
\section{Phase diagram}
The differential equation of the phase variable is completed by the following constitutive choice:
$$
m= \theta +\l  p + \bv_s^2-\bv_n^2,
$$
Hence,  (\ref{eq_vf0}) reads
\begin{equation}\label{eq_vf}
\tau \r \dot \vf = \frac{1}{\k^2} \nabla \cdot (\r\nabla \vf) - \r \theta_\l \, \vf(\vf^2-1) - \r(\theta +\l  p + \bv_s^2-\bv_n^2) \vf.
\end{equation}
As pointed out in Sect.1, in homogeneous and steady conditions the material is in the normal state when $m>\theta_\l$, that is when
\begin{equation}\label{18}
\theta +\lambda  p + \bv_s^2-\bv_n^2 > \theta_\l
\end{equation}
and the transition occurs when $\theta +\lambda  p + \bv_s^2-\bv_n^2 = \theta_\l$.
Therefore the model is naturally able to account for the existence of a critical velocity $\bv_s$ (depending on the temperature and the pressure), above which superfluid properties disappear (see \cite{TT}). 

The regions where $\theta +\lambda  p + \bv_s^2-\bv_n^2$ is respectively grater and smaller than $\theta_\l$ are the stability regions of the normal and superfluid phase and the curve represented by the equation $\theta +\lambda  p + \bv_s^2-\bv_n^2 = \theta_\l$ separates such regions. In particular, if we consider the equilibrium states, {\it i.e.} $\bv_s=\bv_n=0,$ then the curve is represented by the equation
$$
\theta +\lambda p= \theta_\l,
$$ 
which is a line with negative slope $-1/\lambda$. This is good approximation of the $\lambda -$line shown in the phase diagram of liquid helium and represented in Fig.\ref{lam}.

In a previous model for superfluidity (\cite{F_super}) the coefficient $\l$ was supposed to vanish. This corresponds to approximate the $\l-$line with a vertical line. Moreover in such a model the normal component was assumed to be incompressible, so that 
$$
\div\bv_n=0.
$$
In this paper, we consider a different case, by assuming that the normal component obeys the constraint
\begin{equation}\label{eq_p}
	\div\bv_n = \l \r \vf\dot\vf
\end{equation}
which generalizes incompressibility.

\section{Heat equation and thermodynamics}
In order to obtain the kinetic equation for the temperature, let us consider the first law of thermodynamics in the form (\cite{fremond})
\begin{equation}\label{I_law}
	\rho \dot E = \cP^i_\vf + \cP^i_{\bv_s} + \rho h,
\end{equation}
where $E$ is the total energy, $\cP^i_\vf, \cP^i_{\bv_s}$ are respectively the internal powers due to the order parameter and to the velocity $\bv_s$ and $h$ stands for the rate at which the heat is absorbed by the material.
Hereafter we will consider some approximations of our model, valid in a neighborhood of the transition temperature. In particular, in the expression of the time derivative
$$\dot\vf=\partial_t\vf+\bv\cdot\nabla\vf=\partial_t\vf+\bv_n\cdot\nabla\vf+\vf^2(\bv_s-\bv_n)\cdot\nabla\vf$$ we neglect last term, so that $\dot\vf$ assumes the form
$$
\dot\vf=\partial_t\vf+\bv_n\cdot\nabla\vf.
$$
As a consequence
\begin{equation}\label{lemma}
\nabla\dot\vf=(\nabla\vf)^{\cdot}+(\nabla\bv_n)^T\nabla\vf,
\end{equation}
where the superscript $T$ denotes the transpose of a tensor.

Multiplying equation \eqref{eq_vf} by $\dot \vf$ and accounting for \eqref{eq_p} and (\ref{lemma}), we obtain the power balance related to $\vf$, that is
$$
\cP^i_{\vf} = \cP^e_{\vf}
$$
where the internal and external powers are given by
\begin{eqnarray}
\nonumber
\cP^i_{\vf} &=& \r \frac{d}{dt}\left[\frac{1}{2\k^2}|\nabla \vf|^2 +  \theta_\l \,\left(\frac{\vf^4}{4}-\frac{\vf^2}{2}\right) \right] 
+   \r(\theta + \bv_s^2 -\bv_n^2) \vf \dot\vf  
\\
\label{A}
&&
- \bv_n \cdot \nabla p+ \tau \r \dot \vf^2  +\frac{1}{\k^2}\nabla\bv_n \cdot(\nabla\vf\otimes\nabla\vf),
\\
\cP^e_{\vf} &=& \nabla \cdot \left(\frac{1}{\k^2} \r\dot\vf\nabla \vf  - p\bv_n \right).	
\end{eqnarray}

Similarly, by multiplying equation \eqref{eq_vs} by $\dot \bv_s + \nabla \phi_s -\nabla \theta$, we obtain
\bea
&&
|\dot \bv_{s} + \nabla \phi_s - \nabla \theta |^2 + \bv_n \cdot \rot \dot \bv_s
+  \r \vf^2\bv_s \cdot (\dot \bv_s + \nabla \phi_s 
-  \nabla \theta )
\\
&&
= 
-\div (\bv_n \times \dot \bv_s + \bv_n \times \nabla\phi_s - \bv_n \times \nabla \theta).
\eea
We substitute the term $\rot \dot\bv_s$ with equation \eqref{eq_vn} and we take (\ref{eq_phi}) into account. Thus, we obtain
$$
\cP^i_{\bv_s} = \cP^e_{\bv_s}
$$
where the the internal and external powers due to the velocity $\bv_s$ are
\begin{eqnarray}\label{AA}
\cP^i_{\bv_s} & = &
|\dot \bv_{s} + \nabla \phi_s - \nabla \theta |^2 + \frac12 \rho \frac{d}{dt}|\bv_n|^2 -\rho\vf^2\bv_n\cdot\dot\bv_n
+\nu |\div\bv_n|^2
+ \bv_n \cdot \nabla p
\\ \nonumber
&&
-\frac{1}{\kappa^2}\nabla\bv_n\cdot(\nabla\vf\otimes\nabla\vf)+  \r \vf^2\bv_s \cdot \dot \bv_s + \tau \k^2 \r \vf^2 \phi_s^2 - \r \vf^2 (\bv_s -\bv_n) \cdot \nabla \theta
\\
\label{B}
\cP^e_{\bv_s}&=& 
-\div [\bv_n \times \dot \bv_s + \bv_n \times \nabla\phi_s - \bv_n \times \nabla \theta - \nu\bv_n \div\bv_n +\frac{1}{\kappa^2}(\nabla\vf\otimes\nabla\vf) \bv_n
\\ \nonumber
&&
+  \r \vf^2\bv_s \phi_s] + \r \bg \cdot \bv_n.
\end{eqnarray}

It is worth noting that, as a consequence of the boundary conditions, after an integration on the domain $\W$ the only contribution given by the external powers is due to the external source $\bg$.

We assume that the total energy $E$ is written as
\begin{equation}\label{E}
E = \frac{1}{2\kappa^2} |\nabla \vf|^2 + \theta_\l \,\left(\frac{\vf^4}{4}-\frac{\vf^2}{2}\right)+ e_0(\theta) +\frac{1}{2}\vf^2\bv_s^2 +\frac{1}{2}(1-\vf^2) \bv_n^2,
\end{equation}
where $e_0(\theta)$ is a function depending only on the temperature. 
We identify the first three terms of (\ref{E}) with the internal energy
$$
e=\frac{1}{2\kappa^2} |\nabla \vf|^2 + \theta_\l \,\left(\frac{\vf^4}{4}-\frac{\vf^2}{2}\right)+ e_0(\theta).
$$
Last two terms of (\ref{E}) involving the normal and the superfluid components of the velocity define the kinetic energy $T$. Hence
$$
E=e+T.
$$
Substitution of (\ref{A}), (\ref{AA}) and (\ref{E}) into \eqref{I_law} yields
\begin{eqnarray}\nonumber
\r h&=& \r e_0'(\theta) \dot \theta - \r\theta \vf\dot \vf - \tau \r \dot \vf^2 - |\dot \bv_{s} + \nabla \phi_s - \nabla\theta |^2
\\
\label{eq_h}
&&
 - \nu |\div\bv_n|^2 - \tau\k^2\r \vf^2 \phi_s^2 +  \r \vf^2 (\bv_s-\bv_n) \cdot \nabla\theta.
\end{eqnarray}

The heat equation is given by
\begin{equation} \label{heat_eq}
	\r h = -\div \bq + \r r,
\end{equation}
where $\bq$ is the heat flux and $r$ is the heat supply.
In this framework, the heat flux is assumed to satisfy the constitutive equation
\begin{equation} \label{q}
\bq = -k_0(\theta) \nabla \theta -  \r \vf^2 \theta (\bv_s-\bv_n),
\end{equation}
where $k_0(\theta)$ denotes the thermal conductivity.
Notice that, when the fluid is in the normal phase, {\it i.e.} $\vf=0$, equation (\ref{q}) reduces to the usual Fourier law. On the contrary, in the superfluid state, the superfluid component of the velocity $\bv_s$ is related to the heat flux inside the material.

By comparing (\ref{eq_h}) with \eqref{heat_eq}, we obtain the evolution equation for the temperature, {\it i.e.}
\begin{eqnarray}\label{temp}
\r e_0'(\theta) \dot \theta 
&=& \r \theta \vf\dot \vf + \tau \r \dot \vf^2 + |\dot \bv_{s} + \nabla \phi_s - \nabla\theta |^2 + \nu |\div\bv_n|^2 \nonumber
\\
&&
+ \tau\k^2\r \vf^2 \phi_s^2 + \div [k_0(\theta)\nabla \theta] + \div[ \r \vf^2 (\bv_s -\bv_n)] \theta + \r r.
\end{eqnarray}
Equation (\ref{temp}) is completed by the boundary condition
$$
\bq \cdot \bn |_{\partial \Omega} = 0.
$$
In view of the constitutive equation \eqref{q} and the boundary conditions \eqref{bc_vs} and \eqref{bc_vn}, we conclude that
\begin{eqnarray}
\label{b_u}
\nabla \theta \cdot \bn |_{\partial \Omega} = 0.
\end{eqnarray}

Now we prove that our model is consistent with the second law of
thermodynamics.
We write the Clausius-Duhem inequality in the form
$$
\rho \dot \eta \geq -\div\left(\frac{\bq}{\theta} \right) + \frac{\rho r}{\theta},
$$
where $\eta$ is the entropy.
We introduce the Helmholtz free energy density $\Psi = e -\eta \theta$ which is supposed to depend on $(\theta,\vf,\nabla\vf)$.
In view of (\ref{I_law}) and (\ref{heat_eq}), we deduce
$$
\r \dot \Psi + \r \eta \dot\theta +\r \dot T  \leq \cP^i_\vf + \cP^i_{\bv_s} - \frac{1}{\theta} \bq \cdot \nabla \theta.
$$
Accordingly, we have
\bea
\r (\partial_{\theta}\Psi + \eta) \dot \theta + \r [\partial_{\vf}\Psi - \theta_\l \, \vf(\vf^2-1) -\theta \vf ] \dot \vf 
+\r \left[\partial_{\nabla\vf}\Psi - \frac{1}{2\kappa^2}\nabla\vf \right]\cdot \nabla\dot\vf
\\
- \tau \r \dot \vf^2 - |\dot \bv_{s} + \nabla \phi_s - \nabla \theta |^2
-\nu |\div\bv_n|^2 - \tau\k^2 \r \vf^2 \phi_s^2 - \frac{k_0(\theta)}{\theta} |\nabla \theta|^2
\leq 0.
\eea
In virtue of the arbitrariness of $(\dot \theta, \dot\vf, \nabla\dot\vf )$, the functions $\bq, \Psi, \eta$ are compatible with the second law of thermodynamics if $\nu, \tau, k_0$ assume non-negative values, and the free energy $\Psi$  satisfies the following conditions:
$$
\partial_{\theta}\Psi = - \eta,
\quad \partial_{\vf}\Psi = \theta_\l \, \vf(\vf^2-1) +\theta \vf,
\quad 
\partial_{\nabla\vf}\Psi = \frac{1}{2\kappa^2}\nabla\vf.
$$
Hence,
\begin{equation}\label{free_energy}
	\Psi = \theta_\l \, \left(\frac{\vf^4}{4}-\frac{\vf^2}{2}\right) + \frac{1}{2}\theta \vf^2 + \frac{1}{2\kappa^2}|\nabla\vf|^2 + \Psi_0(\theta),
\end{equation}
where $\Psi_0$ depends only on the temperature. From the relation $\Psi = e + \theta \partial_{\theta}\Psi$ it follows that
$$
\Psi_0'(\theta) = \Psi_0(\theta) - e_0(\theta).
$$

Finally, we conclude this section by proving that the passage from the normal phase to the superfluid one is a second-order transition since no latent heat is involved.
Indeed, from \eqref{free_energy}, it follows that the entropy assumes the form
$$
\eta = -\partial_{\theta}\Psi =-\frac{1}{2} \vf^2 - \Psi_0'(\theta)
$$
and the latent heat $L$ is given by
$$
L = \theta [\eta(\theta_\lambda,\vf_0(\theta_\lambda)) - \eta(\theta_\lambda,0)],
$$
where $\vf_0$ is the minimum of the function $W(\vf) = \theta_\lambda F(\vf) + G(\vf)$ and $\vf=\vf_0$, $\vf=0$ characterize the pure phases (see Fig.2). Since $\vf_0=0$ when $\theta=\theta_\lambda$, we have that $L=0$.

%=====================================================================
%======================================================================
%=========================================================================
\section{The differential system and gauge invariance}
Collecting the equations of motion we write the system of equations:
\begin{eqnarray}
\label{eq1}
\tau \r \dot \vf &=& \frac{1}{\k^2} \nabla \cdot (\r\nabla \vf) - \r \theta_\l \, \vf(\vf^2-1) - \r(\theta +\l  p + \bv_s^2-\bv_n^2) \vf.
\\
\label{eq2}
\div[\r \vf^2\bv_s] &=& -\tau \k^2 \r\vf^2\phi_s
\\
\label{eq3}
\dot \bv_{s} &=& - \nabla \phi_s - \rot \bv_n - \r\vf^2 \bv_s + \nabla \theta 
\\
\label{eq4}
\rho(1-\vf^2)\dot\bv_n &=& - \rot \rot \bv_n  + \nu \nabla(\div\bv_n) -\nabla p -\frac{1}{\kappa^2}\div(\nabla\vf\otimes\nabla\vf)
\nonumber
\\
&&
-\r \vf^2 \nabla\theta - \rot[ \r \vf^2\bv_s] + \rho \bg.
\\
\label{eq5}
\div \bv_n &=& \l \r \vf \dot\vf
\\
\label{eq6}
\dot \r &=& -\rho \div[(1-\vf^2)\bv_n+\vf^2 \bv_s]
\\
\nonumber
\r e_0'(\theta) \dot \theta &=& \r \theta \vf \dot\vf + \tau \r \dot \vf^2 + |\dot \bv_{s} + \nabla \phi_s - \nabla\theta |^2 + \nu |\div\bv_n|^2 
\\
\label{eq7}
&&
+ \tau\k^2\r \vf^2 \phi_s^2 + \div [k_0(\theta)\nabla \theta] + \div[\r \vf^2 (\bv_s -\bv_n)] \theta + \r r,
\end{eqnarray}
in the unknowns $\vf, \bv_s, \phi_s, \bv_n, p, \rho, \theta$.

As we have pointed in Section 2, the model we propose is similar to the Ginzburg-Landau model of superconductivity. In order to stress this analogy, following \cite{BFG}, we introduce the transformation:
$$
(\vf, \bv_s, \phi_s)\quad \longleftrightarrow \quad (\psi, \bA, \phi)
$$
where 
\bea
\vf = \psi e^{-i\chi}, \quad \bv_s = \bA - \frac{1}{\kappa}\nabla\chi, \quad \phi_s = \phi + \frac{1}{\kappa} \dot\chi,
\eea
$\chi$ is an arbitrary scalar function and $i$ denotes the imaginary unit.

Our aim is to write equations (\ref{eq1})-(\ref{eq7}) by means of the variables $(\psi, \bA, \phi)$.
Multiplying equation (\ref{eq1}) by $ e^{i\chi}$, we obtain
\begin{eqnarray}
\nonumber
\tau \r \dot \psi &=& \left(i\tau \r \vf \dot \chi + \frac{1}{\k^2} \r \Delta\vf + \frac{1}{\k^2} \nabla\r \cdot \nabla\vf \right) e^{i\chi}- 
\r \psi[\theta_\l (|\psi|^2 -1) + \theta +\l p -\bv_n^2] 
\\
\label{Re}
&&
- \r \psi  \left(|\bA|^2 - \frac2\k \bA \cdot \nabla\chi + \frac{1}{\k^2} |\nabla\chi|^2 \right).
\end{eqnarray}
Dividing (\ref{eq2}) by $\kappa\vf$ and substituting the expressions of $\bv_s$ and $\phi_s$, we deduce
\begin{eqnarray}
\nonumber
\tau \r \vf \dot\chi &=& -\frac{1}{\k}\vf \bA\cdot \nabla\r + \frac1{\k^2} \vf \nabla\r \cdot \nabla\chi - \frac{2}{\k}\r \bA\cdot \nabla\vf 
+ \frac2{\k^2} \r \nabla\vf\cdot \nabla\chi 
\\
\label{Im}
&&
-\frac{1}{\k} \r\vf \div\bA + \frac{1}{\k^2} \r\vf \Delta\chi - \tau\k\r\vf\phi.
\end{eqnarray}
We substitute equation \eqref{Im} into \eqref{Re}. Moreover, the identities
\bea
\nabla \psi &=& (\nabla\vf + i \vf \nabla\chi ) e^{i\chi},
\\
\Delta\psi &=& (\Delta\vf + 2i \nabla\vf \cdot \nabla\chi - \vf|\nabla\chi|^2 + i\vf\Delta\chi) e^{i\chi},
\eea
lead to
\bea
\tau \r \dot \psi &=& \frac{1}{\k^2} \r \Delta\psi + \frac{1}{\k^2} \nabla\psi \cdot \nabla\r - \frac{i}{\k}\psi \bA\cdot \nabla\r
- \frac{2i}{\k}\r \bA\cdot \nabla\psi 
\\
&&-\frac{i}{\k} \r\psi \div\bA -i\tau \k\r\psi\phi 
- \r \psi[\theta_\l(|\psi|^2 -1) + \theta +\l p -\bv_n^2] - \r \psi  |\bA|^2.
\eea

In addition, it is easy to prove the relations
\bea
\dot\bv_s + \nabla \phi_s &=& \dot \bA + \nabla \phi
\\
\vf^2 \bv_s &=& -|\psi|^2 \bA + \frac{i}{2\kappa} (\psi \nabla\psi^* - \psi^*\nabla\psi)
\\
\vf \dot\vf &=& \frac12 (\psi \dot\psi^* + \psi^* \dot\psi)
\\
\dot\vf^2+\k^2\vf^2\phi_s^2&=&|\dot\psi|^2+\k^2\vf^2\phi^2+2\k\phi\vf^2\dot\chi
\\
&=&
|\dot\psi|^2+\k^2 |\psi|^2\phi^2
-i\k\phi(\dot\psi\psi^*-\psi\dot{\psi}^*)
\\
\nabla\vf\otimes\nabla\vf &=&\frac{1}{4|\psi|^2}(\psi^*\nabla\psi+\psi\nabla\psi^*)\otimes(\psi^*\nabla\psi+\psi\nabla\psi^*)
\eea
where $\psi^*$ denotes the conjugate of $\psi$.
Accordingly, the equations (\ref{eq1})-(\ref{eq7}) transform into
\bea
\tau \r \dot \psi &=& \frac{1}{\k^2} \r \Delta\psi - \frac{i}{\k}\psi \bA\cdot \nabla\r + \frac{1}{\k^2} \nabla\psi \cdot \nabla\r
- \frac{2i}{\k}\r \bA\cdot \nabla\psi 
\\
&&-\frac{i}{\k} \r\psi \div\bA -i\tau \k\r\psi\phi 
- \r \psi[\theta_\l (|\psi|^2 -1) + \theta +\l p -\bv_n^2] - \r \psi  |\bA|^2
\\
\dot \bA &=& - \nabla \phi - \rot \bv_n +|\psi|^2 \bA - \frac{i}{2\kappa} (\psi \nabla\psi^* - \psi^*\nabla\psi) + \nabla \theta
\\
\rho(1-|\psi|^2)\dot\bv_n &=& - \rot \rot \bv_n  + \nu \nabla(\div\bv_n)  -\nabla p -\frac{1}{4\k^2}\div\left[\frac{1}{|\psi|^2}(\psi^*\nabla\psi+\psi\nabla\psi^*)\otimes(\psi^*\nabla\psi+\psi\nabla\psi^*)\right]
\\
&&
-\rho|\psi|^2\nabla\theta - \rot\left[-\r|\psi|^2 \bA + \frac{i}{2\kappa} \r(\psi \nabla\psi^* - \psi^*\nabla\psi) \right] + \rho \bg
\\
\div \bv_n &=& \frac{\l}{2} \r (\psi \dot\psi^* + \psi^* \dot\psi)
\\
\dot \r &=& -\rho\div[(1-|\psi|^2)\bv_n-|\psi|^2 \bA + \frac{i}{2\kappa} (\psi \nabla\psi^* - \psi^*\nabla\psi)]
\\
\r e_0'(\theta) \dot \theta &=& \frac12 \r \theta (\psi \dot\psi^* + \psi^* \dot\psi) + \tau \r |\dot \psi|^2 +\tau\r\k^2 |\psi|^2\phi^2
-i\tau\k\r\phi(\dot\psi\psi^*-\psi\dot{\psi}^*) 
\\
&&
+ |\dot \bA + \nabla \phi - \nabla\theta |^2 + \nu |\div\bv_n|^2 + \div [k_0(\theta)\nabla \theta] 
\\
&&
+ \div\left[\r|\psi|^2\bv_n-\r|\psi|^2 \bA + \frac{i}{2\kappa} \r(\psi \nabla\psi^* - \psi^*\nabla\psi) \right] \theta + \r r.
\eea
It is worth noting that the previous equations are independent of $\chi$, which can be chosen arbitrarily.
This formulation allows a more direct comparison with the Ginzburg-Landau model of superconductivity, where the choice of the variables $(\psi,\bA,\phi,\theta)$ is crucial in order to prove well-posedness results for the differential system.


\begin{thebibliography}{99}
\bibitem{BF}  V. Berti and M. Fabrizio,  Existence and uniqueness for a mathematical model in
superfluidity, Math. Meth. Appl. Sci. 31 (2008) 1441--1459.
%%%%%%%%%%%%%%%%%%%%%%%%%%%%
\bibitem{BFG}  V. Berti, M. Fabrizio and C. Giorgi, Gauge invariance and asymptotic behavior for
the Ginzburg-Landau equations of superconductivity, J. Math. Anal. Appl. 329 (2007) 357-–375.
%%%%%%%%%%%%%%%%%%%%%%%%%%%%%%%
\bibitem{Capriz} G. Capriz, Continua with Microstructure, Springer-Verlag, New York, 1989.
%%%%%%%%%%%%%%%%%%%%%%%%%%%%%%%%
\bibitem{PPG} G. Capriz and P. Podio-Guidugli, Materials with spherical structure, Arch. Rational Mech.
Anal. 75 (1981) 269--279.
%%%%%%%%%%%%%%%%%%%%%%%%%%%%%
\bibitem{Fab06}  M. Fabrizio, Ginzburg-Landau equations and first and second order phase transitions, Internat. J. Engrg. Sci. 44 (2006) 529--539.
%%%%%%%%%%%%%%%%%%%%%%%%%%%%
\bibitem{F_super}   M. Fabrizio, A Ginzburg-Landau model for the phase transition in Helium II, Z.Angew.Math.Phys. 61 (2010) 329--340.
%%%%%%%%%%%%%%%%%%%%%%%%%%%
\bibitem{FM}  M. Fabrizio and A. Morro, Electromagnetism of Continuous Media. Oxford University Press, Oxford 2003.
%%%%%%%%%%%%%%%%%%%%%%%%%%%
\bibitem{fremond}  M. Fr\'emond, Non-smooth Thermomechanics. Springer, Berlin (2002).
%%%%%%%%%%%%%%%%%%%%%%%%%%%
\bibitem{GL}  V.L.~Ginzburg and L.D.~Landau, On the theory of superconductivity, Zh. Eksp. Teor. Fiz. 20 (1950) 1064--1082.
%%%%%%%%%%%%%%%%%%%%%%%%%%%%%%%%%%%%%%%%%%%%%%%%%%%%%%%%%%
\bibitem{G}
E. Fried and M. E. Gurtin, Continuum theory of thermally induced phase transitions based on
an order parameter, Physica D 68 (1993) 326--343.
%%%%%%%%%%%%%%%%%%%%%%%%%%%%%%%%%%%%%%
\bibitem{fried} E. Fried and M. E. Gurtin, Dynamic solid-solid transitions with phase characterized by an order
parameter, Physica D 72 (1994) 287--308.
%%%%%%%%%%%%%%%%%%%%%%%%%%%%%%%%%%%%%%%
\bibitem{landau} L.D. Landau, On the theory of superfluidity of helium II, Journal of Physics USSR 5 (1941) 71--77.
%%%%%%%%%%%%%%%%%%%%%%%%%%%%%%%%%%%%
\bibitem{LH}   L.~Lindblom and W.~Hiscock, A one-fluid model of superfluids, Physics Letters A, 131 (1988) 280--284.
%%%%%%%%%%%%%%%%%%%%%%%%%%%%%%%
\bibitem{Mend} K. Mendelssohn, Liquid Helium. In Handbuch Physik, Flugge S, Springer, Berlin 1956, 370–461.
%%%%%%%%%%%%%%%%%%%%%%%%%%%%%%%%%%%%%
\bibitem{M}  M. S.~Mongiov\'i, Extended  irreversible thermodynamics of liquid helium II, Phys. Rev. B, 48 (1993) 6276--6283.
%%%%%%%%%%%%%%%%%%%%%%%%%%
\bibitem{TT} D.R. Tilley and J.Tilley,  Superfluidity and superconductivity, Graduate student series in physics 138, Bristol 1990.
%%%%%%%%%%%%%%%%%%%%%%%%%%%%%
\bibitem{Tisza} L. Tisza, Transport phenomena in He II, Nature, 141 (1938) 913.
%%%%%%%%%%%%%%%%%%%%%%%%%%%%%
\bibitem{T}  {M. Tinkham},  Introduction to superconductivity. McGraw-Hill, New York 1975.
%%%%%%%%%%%%%%%%%%%%%%%%%%%%
\end{thebibliography}
\end{document}